\documentclass{moriond}

\def\be{\begin{equation}}
\def\ee{\end{equation}}
\def\bea{\begin{eqnarray}}
\def\eea{\end{eqnarray}}

\newcommand{\Photo}{}

\usepackage[normalem]{ulem}

\begin{document}
\vspace*{4cm}
\title{Binary systems as  gravitational wave detectors}

\author{Diego Blas\footnote{Presenter at 2025 Gravitation session of the 59th Rencontres de Moriond. Email address: \href{mailto:dblas@ifae.es}{dblas@ifae.es}.}}

\address{Institut de Fisica d’Altes Energies (IFAE), The Barcelona Institute of Science and Technology, Campus UAB, 08193 Bellaterra (Barcelona), Spain, and \\ 
Instituci\'o Catalana de Recerca i Estudis Avan\c cats (ICREA), Passeig Llu\'is Companys 23, 08010 Barcelona, Spain}

\author{Adrien~Bourgoin}
\address{LTE, Observatoire de Paris, Universit\'e PSL, Sorbonne Universit\'e, Universit\'e de Lille, LNE, CNRS 61 Avenue de l'Observatoire, 75014 Paris, France}

\author{Joshua~W.~Foster}
\address{Astrophysics Theory Department, Theory Division, Fermilab, Batavia, IL 60510, USA}

\author{Aurelien~Hees}
\address{LTE, Observatoire de Paris, Universit\'e PSL, Sorbonne Universit\'e, Universit\'e de Lille, LNE, CNRS 61 Avenue de l'Observatoire, 75014 Paris, France}
\author{M\'iriam~Herrero-Valea}
\address{Barcelona Supercomputing Center (BSC), Plaça d'Eusebi Güell 1-3, 08034 Barcelona, Spain}
\author{Alexander~C.~Jenkins}
\address{Kavli Institute for Cosmology, University of Cambridge, Madingley Road, Cambridge CB3 0HA, UK\\
DAMTP, University of Cambridge, Wilberforce Road, Cambridge CB3 0WA, UK}
\author{Xiao~Xue}
\address{Institut de F\'{i}sica d’Altes Energies (IFAE), The Barcelona Institute of Science and Technology,
Campus UAB, 08193 Bellaterra (Barcelona), Spain}

\maketitle\abstracts{``Adding colour'' to the gravitational-wave sky by probing unexplored frequency bands has the potential to unveil new and unexpected phenomena in the Universe. In these proceedings, we summarise current efforts to probe gravitational waves with $\mu$Hz frequencies through their resonant impact on binary systems. We also discuss how these binaries can explore uncharted regions of parameter space in the hunt for dark matter.}

\section{Introduction}

Gravitational waves (GWs) have established themselves as an exciting new messenger of the Cosmos. As compared to light signals, GWs are produced by all energetic sources with non-trivial dynamics and can propagate through the primordial Universe without being scattered or absorbed. This explains why their first detection in 2015 was so consequential: a new ``dark'' universe, completely hidden from our previous telescopes (mainly triggered by light), started to manifest itself. The impact on gravitational physics, astrophysics, fundamental physics, cosmology, or particle physics of this and subsequent discoveries is hard to overestimate. 

However, direct detection has only been achieved at frequencies around 100 Hz \cite{LIGOScientific:2016aoc} and, more recently, strongly hinted in the nHz band \cite{Antoniadis:2023rey,Reardon:2023gzh,NANOGrav:2023gor,Xu:2023wog}. The study of ways to discover GWs in a broader frequency band, to ``add colour" to the gravitational wave sky, will in the following years extend the relevance of GWs to completely uncharted territories, where new phenomena from the dark Universe reside. In this contribution, we will show how the orbital dynamics of binary systems are impacted by GWs and how we can use measurements from binary systems to unveil GWs in a frequency band that is currently unexplored.

Furthermore, the dark sector of the Universe (dark matter and dark energy), whose mysteries are still far from being unveiled, may also generate fluctuations in the gravitational potentials that determine the dynamics of these binaries. These effects, as we will see, also generate modifications in the orbital motion of these systems that can be sensitive to models that are currently inaccessible.     

We work in units where $\hbar=c=1$.

\section{Gaps in the gravitational wave spectrum}

Gravitational-wave signals can either be described in terms of a waveform for each transverse-traceless polarisation (``$+$'' and ``$\times$''), typically in the frequency domain, $h_{+,\times}(f)$, or, in cases where the phase evolution is unknown (such as stochastic backgrounds), in terms of an equivalent, polarisation-averaged ``characteristic strain'' $h_c(f)$ defined by the power spectral density of the signal~\cite{Maggiore:2007ulw}. The latter is related to the cosmological energy density in GWs, $\Omega(f)=\frac{1}{\rho_c}\frac{\mathrm{d}\rho_{\rm GW}}{\mathrm{d}\log(f)}= \frac{2h_{c}^2(f)}{3} \left(\frac{\pi f}{H_0} \right)^2$, expressed as a fraction of the present-day critical density $\rho_c=3H_0^2/(8\pi G)$, with $H_0$ the present-day Hubble rate.

\begin{figure}
    \centering
    \includegraphics[width=0.7\linewidth]{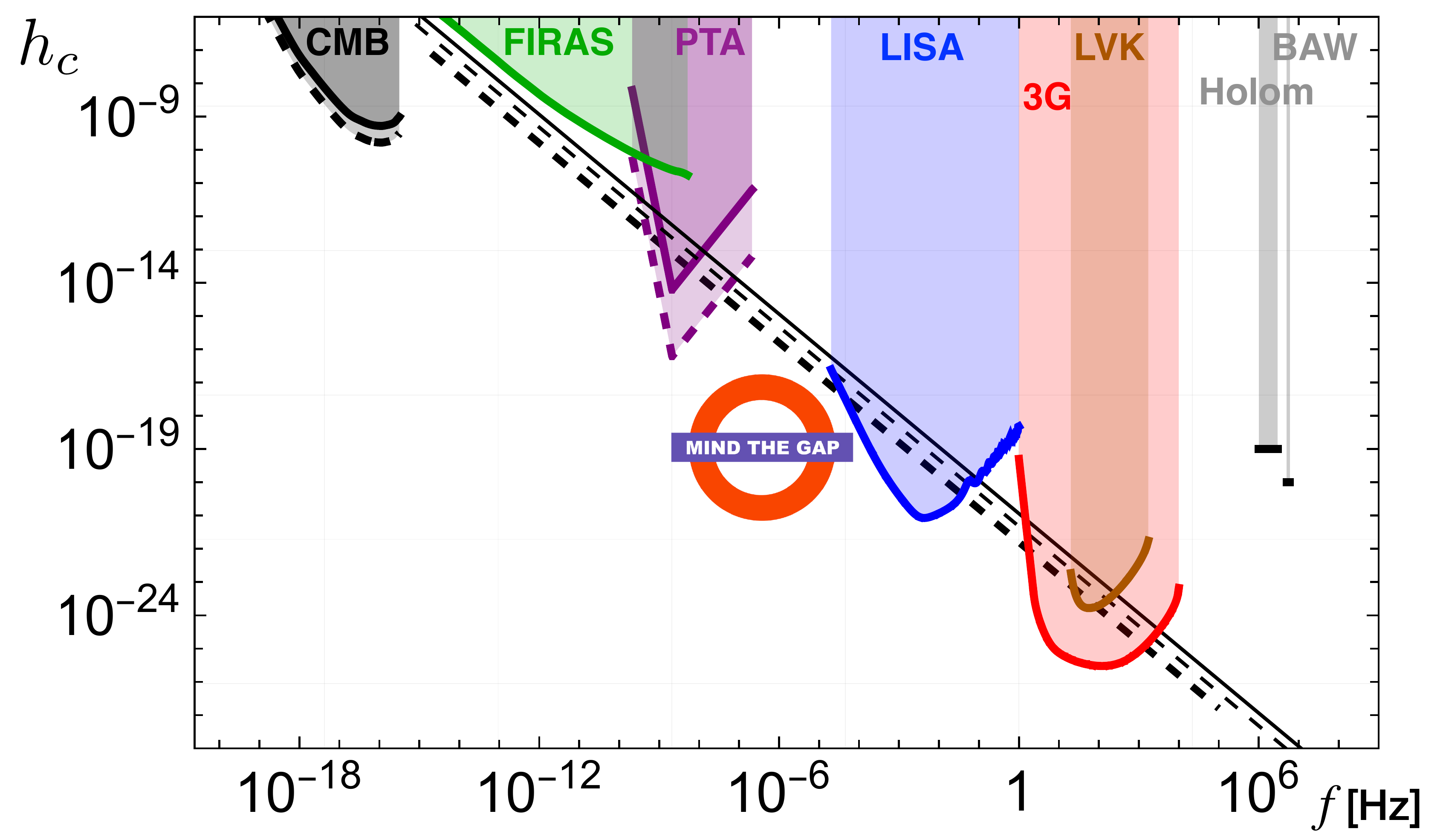}
    \caption[]{Landscape of bounds on the gravitational wave spectrum by 2040. CMB bounds from~\cite{Clarke:2020bil}, FIRAS bound from~\cite{Kite:2020uix}, PTA bounds from~\cite{Lasky:2015lej}, LISA line from~\cite{LISA:2024hlh}, LVK line from~\cite{KAGRA:2021kbb}, 3G from \cite{Branchesi:2023mws}. Holom constraint from~\cite{2017PhRvD..95f3002C}. BAW constraint from~\cite{Goryachev:2021zzn}. The gap in the $\mu$Hz band is highlighted.}
    \label{fig:allbounds}
\end{figure}

The current sensitivity to $h_c(f)$ is shown in Fig.~\ref{fig:allbounds} for various current searches, and for key planned searches expected by 2040. As one can see, there are several gaps left to explore: in the very low-frequency regime there is a gap between the frequencies probed by pulsar timing arrays (PTAs) and the polarisation of the cosmic microwave background (CMB); another gap at $\mu$Hz frequencies appears between PTAs and the planned space-based interferometer LISA; and finally, a large gap appears at high frequencies beyond the reach of ground-based interferometers such as LIGO/Virgo/KAGRA. Each of these represents a new opportunity for detecting different phenomena, much like how the different bands in the electromagnetic spectrum give complementary information about the Cosmos.  There were very nice presentations at 2025 Gravitation session of the 59th Rencontres de Moriond for ideas to cover some of these bands. In this contribution, we will focus on the $\mu$Hz gap, corresponding to timescales of hours to years. 

\subsection{The $\mu$Hz gap: a band full of sources}

The $\mu$Hz gap has a particular relevance for GW searches for several reasons. First, it corresponds to a band where we expect to have signals at the level of $h_c\sim 10^{-17}$ from a plethora of compact binaries across a huge range of masses~\cite{Sesana:2019vho}.  As part of this, if the strong evidence of a stochastic gravitational wave background (SGWB) found in PTA searches at nHz frequencies~\cite{Antoniadis:2023rey,Reardon:2023gzh,NANOGrav:2023gor,Xu:2023wog} corresponds to the signal from a population of supermassive black hole binaries, we expect this signal to have support in the $\mu$Hz band at a similar characteristic amplitude, see e.g.~\cite{Ellis:2023owy}. Furthermore, signals from fundamental Physics (from inflation, cosmic strings, first-order phase transitions, or from the whole population of binaries generated by primordial black holes) may also populate this band at characteristic amplitudes that may be even stronger, which may allow us to distinguish between a signal from an astrophysical population and that of an early Universe event. This provides strong motivation for realistic detectors to cover this band.  For the different ideas suggested so far, the interested reader may consult~\cite{Blas:2021mqw,Foster:2025nzf}.

\section{Resonant absorption of gravitational waves in binaries}

\begin{figure}
    \centering
    \includegraphics[width=0.5\linewidth]{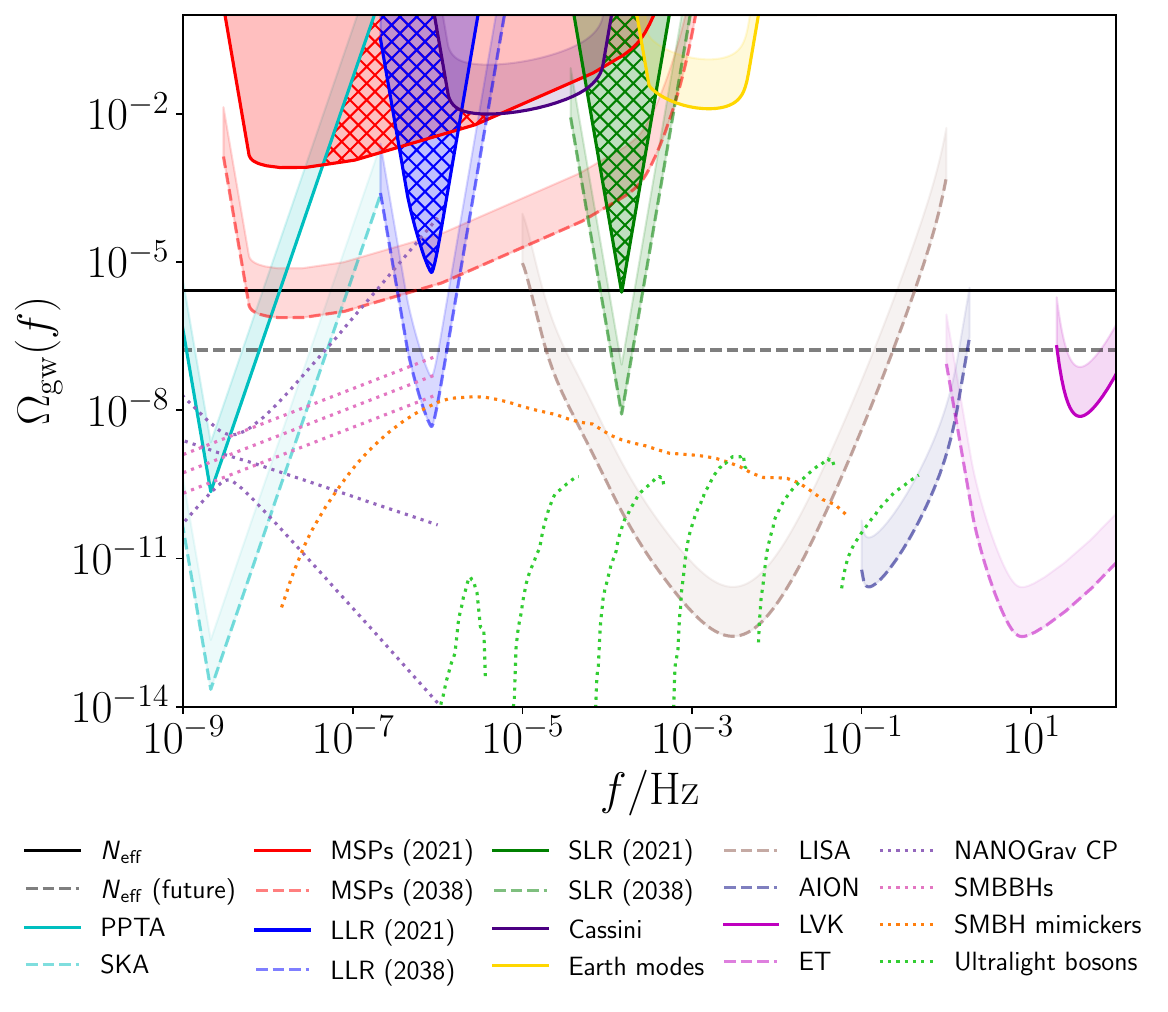}
    \caption[]{Prospects for constraints on $\Omega(f)$ from different binary systems as computed in~\cite{Blas:2021mqw,Blas:2021mpc}, compared to other detection technologies expected by 2035 (see Fig.~\ref{fig:allbounds} for references). We focused on LLR, SLR, and binary millisecond pulsars (MSPs). The information for the MSPs we used is found in~\cite{Blas:2021mqw}. Solid lines of these techniques correspond to what may be achieved in 2021, while dashed ones correspond to what may be achieved in 2035. See main text.  We also showed possible signals in this band. Of particular interest are the lines marked as NanoGrav CP, corresponding to the strong evidence of GWs at PTA searches in the nHz \cite{Antoniadis:2023rey,Reardon:2023gzh,NANOGrav:2023gor,Xu:2023wog}, extended to the $\mu$Hz band in a model agnostic way.}
    \label{fig:2021}
\end{figure}

GWs are searched for in interferometers by very precisely tracking deviations in the trajectories of light (in laser interferometers) or atomic matter waves (in atom interferometry). These effects can be measured with high precision, but do not accumulate in time. On the contrary, searches in \textit{resonant systems} may rely on less sensitive detectors but target effects that do accumulate with time. The quintessential example is a resonant bar, searching for the accumulation of vibrational modes generated by GWs (typically around kHz frequencies). Resonant absorption requires the matching of the frequency of the GW to the typical ones of the resonator. It was suggested in~\cite{Hui:2012yp} that gravitationally-bound binary systems are also natural resonant absorbers of GWs. As in the case of resonant bars, binary systems respond to oscillating GWs by undergoing dynamical changes that accumulate with time when the frequency of the GW satisfies $f=n/P$, with $P$ being the orbital period of the binary, and $n$ a positive integer. For nan SGWB, this effect generates a random walk in the orbital parameters of the binary, resulting in an evolution of their variances over time. All this is based on the simple equation modifying Newton's law of gravitation in the presence of a gravitational wave of wavelength much larger than the size of the binary,
\begin{equation}
  \frac{\mathrm{d}^2 r^{i}}{\mathrm{d} t^2} + \frac{GM}{r^3} r^i = \frac{1}{2} \delta^{il}\frac{\mathrm{d}^2{h}_{lj}}{\mathrm{d} t^2} r^j, \label{eq:acc}
\end{equation}
    where $M$ is the total mass of the system, and $r^i$ is the separation vector between the two objects. This equation can be modified to include all the effects relevant to the binary, such as post-Newtonian effects, third-body interactions, etc. An important remark is that if the orbit is eccentric, $r^i$ evolves periodically in time, and this effect may resonate with the fluctuations of $h_{ij}$ in time, generating a secular force on the orbital motion. One can proceed now to find the effects of this new acceleration on the system. This is most conveniently done in terms of orbital parameters, which are either constant or trivially evolving in time for a Newtonian binary. These are the semilatus rectum, the eccentricity, the inclination, the longitude of the ascending node, the argument of periapsis, and the true anomaly. The evolution of all these parameters for an SGWB was first derived in~\cite{Blas:2021mpc,Blas:2021mqw}. These works concluded that current data of the evolution of the period and its covariance from binary pulsars, lunar laser ranging (LLR), and satellite laser ranging (SLR) seemed to be on the verge of covering the interesting $\mu$Hz gap with unique precision, with the possibility of accessing the relevant $h_c(f)\sim 10^{-17}$ ($\Omega(f)\sim 10^{-8}$) we discussed above in the near future. These constraints are shown in Fig.~\ref{fig:2021}, where the solid lines correspond to what may be achieved today, while the dashed correspond to what could be achieved by 2035. The bounds (except for Holom and BAW, which correspond to monochromatic waves) are calculated for integrating power spectra, as is usual in searches of SGWBs, see~\cite{Foster:2025csl}.

\section{Beyond-secular modeling of gravitational-wave resonance in binaries}

\begin{figure}
    \centering
    \includegraphics[width=0.8\linewidth]{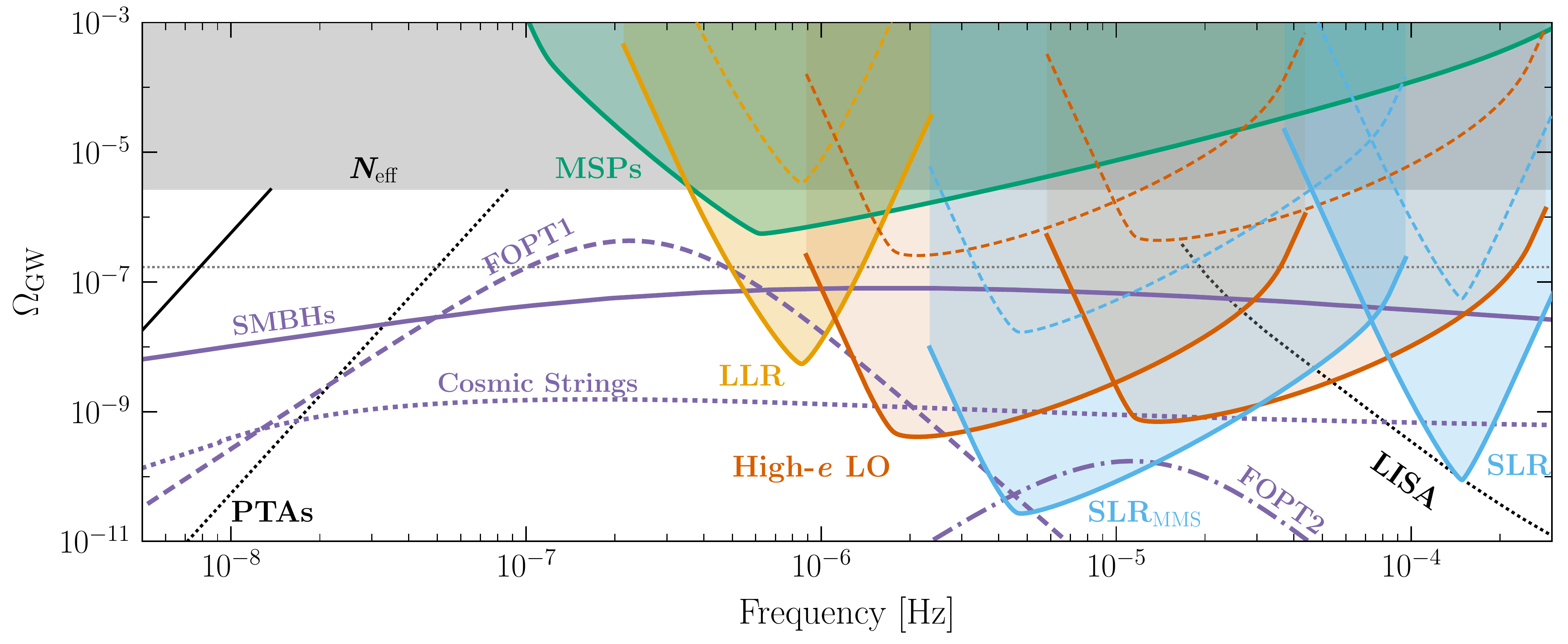}
    \caption[]{Prospects for constraints on $\Omega(f)$ from different binary systems as computed in~\cite{Foster:2025csl,Foster:2025nzf}, compared to other detection technologies expected by 2040 (see Fig.~\ref{fig:allbounds} for references). We focused on LLR, SLR, and  MSPs (see~\cite{Foster:2025nzf} for the MSPs we used). For SLR we show the line corresponding to LAGEOS (SLR line), a possible Earth-bound mission with orbit following that of MMS mission (SLR${}_{\rm MMS}$ line) and two possibilities for Moon-bound orbits (High-e LO), representative of NASA Gateway orbit \url{https://www.nasa.gov/mission/gateway/} and ESA Moonlight constellation \url{https://www.esa.int/Applications/Connectivity_and_Secure_Communications/ESA_s_Moonlight_programme_Pioneering_the_path_for_lunar_exploration}. Solid lines are the optimistic reach, while dashed lines correspond to a dramatic lost of sensitivity in the pessimistic case where a large degeneracy with other parameters is present, see~\cite{Foster:2025nzf}. We also show possible signals in this band: cosmic strings and FOPT1 and FOPT2 represent possible signals from the early universe, while SMBHs corresponds to the strong evidence of GWs at PTA searches in the nHz \cite{Antoniadis:2023rey,Reardon:2023gzh,NANOGrav:2023gor,Xu:2023wog}, extended to the $\mu$Hz band, assuming that the signal comes from supermassive black hole binaries.}
    \label{fig:2025}
\end{figure}

The results of~\cite{Blas:2021mpc,Blas:2021mqw} were derived assuming a perfect non-degeneracy between the effect from the GWs and those usually studied when precise orbit determination is performed, or that our predictions do not deteriorate when the data is analysed. These seem quite optimistic assumptions, and our aim in the last year has been to bridge from the theoretical predictions to a realistic data analysis. In~\cite{Foster:2025csl,Foster:2025nzf}, we have performed a more systematic study of resonant binaries, this time in terms of the real observables in laser ranging missions (residuals in the time of arrival of two-way laser shot from Earth, reflected at the satellite, and received later on at Earth) and in pulsar timing analysis (residuals in the time of arrivals of radio pulses from millisecond pulsars). The most relevant conclusion of these works is that by using \textit{all the information available} in the system (working not only with the secular effects on the period) \textit{the sensitivity to GWs grows dramatically}. Furthermore, it has a more favourable scaling with time of the effects of GWs ($\propto t$ for the period vs. $\propto t^2$ for the changes in the true anomaly, where $t$ is the time variable). Taking this into account, we have shown that the bounds that can be achieved by LLR, SLR, and pulsar timing with data from 2025 (if no degeneracies or other problems are present) are those shown in Fig.~\ref{fig:2025}. They represent an improvement of almost two orders of magnitude! 

This plot contains three forecasts for satellite laser ranging: SLR, SLR${}_{\rm MMS}$, and High-e LO. SLR corresponds to orbits and data that is available today (LAGEOS-1 satellite) if no major degeneracy with the effect of GWs is found in the analysis. However, this orbit (and indeed most current SLR orbits) is not optimal to search for GWs for two main reasons: first, it is in the range of frequencies that LISA will cover with much better sensitivity; second, the orbit is almost circular, and therefore does not efficiently accumulate the effect from GWs. SLR${}_{\rm MMS}$ corresponds to what seems possible to achieve in a dedicated mission of a single satellite at orbits close to those of the NASA mission MMS, see \url{https://mms.gsfc.nasa.gov/}. The choice of this orbit is to demonstrate which orbits could make the SLR concept around Earth achieve sensitivities with real discovery potential. These orbits present their own challenges, such as large aberration at perigee, very distant apogee, peculiar acceleration noises, etc. that require a dedicated study. We expect to provide some conclusions in this direction in the near future. A first estimate shows that such a mission could be delivered with a budget of a few hundred million Euros. We expect that this moderate cost will motivate space agencies to further study their viability. Finally, an SLR mission with an eccentric orbit of a period larger than days would extend the searches of new forces and also help in other studies in geodesy, see e.g. \cite{2014PhRvD..89h2002L,Lucchesi:2010zzb}. 

\section{Beyond-secular modeling of ultra-light dark matter resonance in binaries}

In previous sections, we studied the impact of the spacetime metric fluctuations characteristic of GWs on orbital motion. Another source of metric fluctuations that may affect this motion is the gravitational field of the dark matter halo of the Galaxy, specifically in scenarios where dark matter is ultra-light (with mass $m_{\rm DM}\ll$\,eV). These dark matter models feature a collection of non-relativistic bosons with a typical separation smaller than their de Broglie wavelength. As a result, they are most efficiently described as a stochastic classical field with large regions where the dark matter field oscillates coherently (with large coherence times) at frequency $\omega=m_{\rm DM}$, while the associated gravitational potential felt by the binary systems ($\psi$, generating an acceleration $a^i=-\ddot\psi \,r^i$ in the right-hand side of Eq.~(\ref{eq:acc})) oscillates at $\omega=2m_{\rm DM}$. This oscillation has a very narrow width ($\Delta \omega/\omega\sim 10^{-6}$). The same field configuration also generates a stochastic background of oscillations of $\psi$ with support from  $\omega\approx10^{-6}m_{\rm DM}$ to $\omega=0$. 

\begin{figure}
    \centering
    \includegraphics[width=0.6\linewidth]{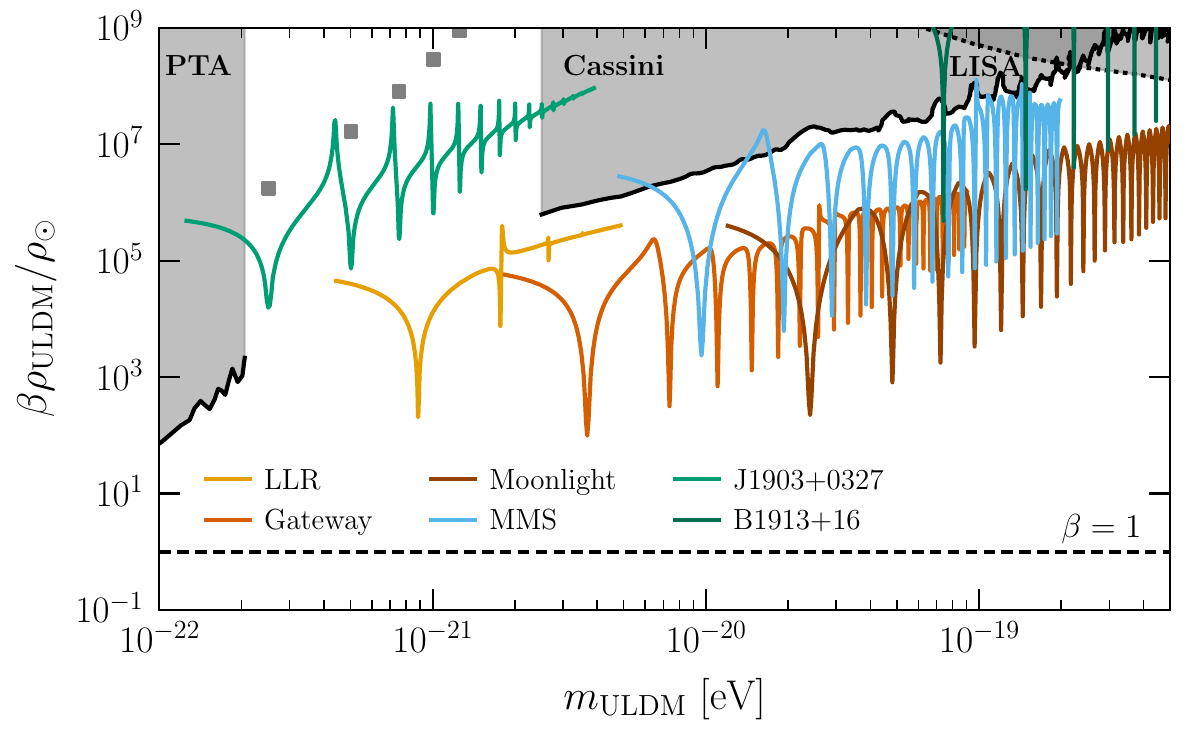}
    \caption[]{Projections for the sensitivities of the binary resonance scenarios considered in this work to ultra-light dark matter. Grey squares represent the on-resonance sensitivity for J1903+0327 derived in \cite{Blas:2016ddr}. Cassini plots follow from ~\cite{Bertotti:2003rm}. Data for J1903+0327 from~\cite{Champion:2008ge,2011MNRAS.412.2763F} and B1913+16 from~\cite{Weisberg:2004hi}. See text and previous figures for more details.}
    \label{fig:ULDM}
\end{figure}

The idea to search for these oscillations in pulsar timing was first explored in~\cite{Blas:2016ddr,Blas:2019hxz}, which focused on the resonant effect of the oscillations of the dark matter field and the corresponding gravitational field on the period and eccentricity of binary pulsars. In~\cite{Foster:2025csl,Foster:2025nzf}, the focus has instead been the time evolution of the true anomaly, and the observational applications have been extended to LLR and SLR. As in the case of GWs, this leads to much stronger constraints than in previous work. These studies conclude that even if current and future missions will not be sensitive to the expected values of $\psi$ from the Galactic dark matter, they are already able to probe new parameter space related to the coupling of dark matter with ordinary matter. In Fig.~\ref{fig:ULDM}, we show projections for the sensitivity to the local density of ULDM, normalized to the expected one in the Solar system, $\rho_\odot$, through searches for the effects of the oscillations it induces in a local potential experienced by binary systems. The parameter $\beta$ represents a possible direct coupling of dark matter to standard model matter. A coupling larger than the gravitational one (normalized to $\beta=1)$ can be converted into a larger effective source of the effect, which in some models simply means that the effective density experienced by the binary is larger (represented by a large $\beta$). Remarkably, the whole parameter space which is not constrained in Fig.~\ref{fig:ULDM} corresponds to possible dark matter models that are currently undetected. We hence see how pulsars, LLR, and SLR may be used to achieve a local detection of dark matter in the ultra-light regime. In the same plot, we have used squares to mark the peak of sensitivity claimed in~\cite{Blas:2016ddr} for J1903+0327. This allows us to show how all the data available from orbital motion may be used to obtain much better sensitivity than the use of the evolution of the period that was studied in~\cite{Blas:2021mqw} (one needs to compare only the peaks of sensitivity to see the improvement).

\section{Conclusions}

The exploration of detection techniques of GWs in the $\mu$Hz band will most likely have consequential results for our understanding of gravitational phenomena, such as the dynamics of massive and supermassive black holes and the physics of the early Universe~\cite{Sesana:2019vho}. Achieving the sensitivity behind this claim seems challenging for interferometers. When thinking about other techniques, resonant absorption by binaries stands out, as the required characteristic resonant periods correspond to hours to months, timescales which occur naturally in binary pulsars and satellite orbits close to the Earth (including that of the Moon).
Since these systems have been tracked with exquisite precision since the 1970s, and resonant effects accumulate with time, one may hope that existing and forthcoming data may already be able to achieve substantial discoveries. This idea was explored in~\cite{Hui:2012yp,Blas:2021mpc,Blas:2021mqw}, with promising forecasts using the modification in the orbital period by a stochastic background of GWs with support in the $\mu$Hz band ($\Omega_{\mathrm{gw}} \geq 4.8 \times 10^{-9}$ for $f=0.85\, \mu\mathrm{Hz}$, corresponding to the lunar period, with future LLR data, cf.~Fig.~\ref{fig:2021}). We have recently revised these forecasts in~\cite{Foster:2025csl,Foster:2025nzf} and found that the inclusion in the parameter estimation of all the available data (in particular the evolution of the binary on sub-orbital timescales) drastically improves this sensitivity. As shown in Fig.~\ref{fig:2025}, even with today's LLR data one may achieve sensitivity to the SGWB that is an order of magnitude stronger than the one previously expected for 2035, cf.~Fig.~\ref{fig:2025}. This is based on the hypothesis that no degeneracy with another parameter of the LLR analysis deteriorates our sensitivity. We have already shown in~\cite{Foster:2025csl} that the tidal deformation of the Earth from the Moon perturbs the Moon's orbit in a way that may be degenerate with certain GW signals. We therefore plan to perform a more detailed analysis to convert the ``expectations'' in Fig.~\ref{fig:2025} into real bounds. This analysis is underway.

Our results since 2021 also show that SLR is a promising technique to bridge the frequencies between LISA and LLR. In~\cite{Blas:2021mpc} we only considered data from existing SLR missions. In~\cite{Foster:2025nzf}, we have extended this analysis, not only to enhance our forecast constraints by using a better data analysis strategy but also by suggesting other missions that, if laser-ranged with currently-achievable accuracy, would most likely detect GWs from supermassive black holes, cf.~Fig.~\ref{fig:2025}. We find these prospects very exciting, and worth a dedicated (and relatively cheap) mission. 

Binary millisecond pulsars can provide wider coverage of GWs, as one can use several of them as local resonant probes of gravitational perturbations, cf.  Fig.~\ref{fig:2025}. We are currently working towards a realistic analysis of data from some selected binaries. The analysis of the orbits from other satellites, such as those in GNSS constellations, or in lunar orbits, may also be relevant to detect GWs. The former is discussed in the contribution to these proceedings by S. Roy~\cite{Roy:2025iug}.

Binary orbits are also affected by the fluctuations from ultra-light dark matter. In Fig.~\ref{fig:ULDM} we have plotted the reach of six binary systems to the dark matter energy density, normalized by the expected dark matter energy density at the Sun's location, and as a function of the mass of the dark matter candidate. We also include a parameter $\beta$ that encapsulates the possibility of a stronger interaction between matter and dark matter ($\beta=1$ corresponds to the universal coupling and larger values to other possible interactions). We see how much parameter space can be won by simply using these six systems. As for the case of GWs, a more detailed analysis is in progress to confirm these predictions. 

To summarize, it seems that binary resonances in binary pulsars and satellite orbits are an efficient way to search for $\mu$Hz GWs and ultra-light dark matter. More precise studies are required to validate this claim. The sensitivity of these searches is likely to be limited not by statistics but by systematic uncertainties due to non-gravitational nuisance accelerations acting on the binary. While we have not discussed this in this short contribution, we believe that these systematics can be mitigated by cross-correlating data from multiple satellites with large eccentricities and long periods. We hope that in the future a dedicated satellite mission with at least two spacecraft will allow us to confirm our forecasts, and give more colour to the GW and dark matter sky.

\section*{Acknowledgments}

D.B. acknowledges the support from the Departament de Recerca i Universitats from Generalitat de Catalunya to the Grup de Recerca 00649 (Codi: 2021 SGR 00649).
The research leading to these results has received funding from the Spanish Ministry of Science and Innovation (PID2020-115845GB-I00/AEI/10.13039/501100011033). This publication is part of the grant PID2023-146686NB-C31 funded by MICIU/AEI/10.13039/501100011033/ and by FEDER, UE.
IFAE is partially funded by the CERCA program of the Generalitat de Catalunya.
Project supported by a 2024 Leonardo Grant for Scientific Research and Cultural Creation from the BBVA Foundation. The BBVA Foundation accepts no responsibility for the opinions, statements and contents included in the project and/or the results thereof, which are entirely the responsibility of the authors. 
This work is supported by ERC grant ERC-2024-SYG 101167211. Funded by the European Union. Views and opinions expressed are however those of the author(s) only and do not necessarily reflect those of the European Union or the European Research Council Executive Agency. Neither the European Union nor the granting authority can be held responsible for them.
D.B. acknowledges the support from the European Research Area (ERA) via the UNDARK project of the Widening participation
and spreading excellence programme (project number 101159929). 
A.C.J. was supported by the Engineering and Physical Sciences Research Council [grant number EP/U536684/1], and by the Gavin Boyle Fellowship at the Kavli Institute for Cosmology, Cambridge.
This research used resources from the Lawrencium computational cluster provided by the IT Division at the Lawrence Berkeley National Laboratory, supported by the Director, Office of Science, and Office of Basic Energy Sciences, of the U.S. Department of Energy under Contract No. DE-AC02-05CH11231.
Part of this work was carried out at the Munich Institute for Astro-, Particle and BioPhysics (MIAPbP), which is funded by the Deutsche Forschungsgemeinschaft (DFG, German Research Foundation) under Germany's Excellence Strategy – EXC-2094 – 390783311. This manuscript has been authored in part by Fermi Forward Discovery Group, LLC under Contract No. 89243024CSC000002 with the U.S. Department of Energy, Office of Science, Office of High Energy Physics.
The work of M. H-V. has been partially supported by the Spanish State Research Agency MCIN/AEI/10.13039/501100011033 and the EU NextGenerationEU/PRTR funds, under grant IJC2020-045126-I.
X.X. is funded by the grant CNS2023-143767. 
Grant CNS2023-143767 funded by MICIU/AEI/10.13039/501100011033 and by European Union NextGenerationEU/PRTR.

\section*{References}
\bibliography{moriond}

\begin{thebibliography}{10}

\bibitem{LIGOScientific:2016aoc}
B.~P. Abbott et~al.
\newblock {Observation of Gravitational Waves from a Binary Black Hole Merger}.
\newblock {\em Phys. Rev. Lett.}, 116(6):061102, 2016.

\bibitem{Antoniadis:2023rey}
J.~Antoniadis et~al.
\newblock {The second data release from the European Pulsar Timing Array III. Search for gravitational wave signals}.
\newblock 6 2023.

\bibitem{Reardon:2023gzh}
Daniel~J. Reardon et~al.
\newblock {Search for an Isotropic Gravitational-wave Background with the Parkes Pulsar Timing Array}.
\newblock {\em Astrophys. J. Lett.}, 951(1):L6, 2023.

\bibitem{NANOGrav:2023gor}
Gabriella Agazie et~al.
\newblock {The NANOGrav 15 yr Data Set: Evidence for a Gravitational-wave Background}.
\newblock {\em Astrophys. J. Lett.}, 951(1):L8, 2023.

\bibitem{Xu:2023wog}
Heng Xu et~al.
\newblock {Searching for the Nano-Hertz Stochastic Gravitational Wave Background with the Chinese Pulsar Timing Array Data Release I}.
\newblock {\em Res. Astron. Astrophys.}, 23(7):075024, 2023.

\bibitem{Maggiore:2007ulw}
Michele Maggiore.
\newblock {\em {Gravitational Waves. Vol. 1: Theory and Experiments}}.
\newblock Oxford University Press, 2007.

\bibitem{Clarke:2020bil}
Thomas~J. Clarke, Edmund~J. Copeland, and Adam Moss.
\newblock {Constraints on primordial gravitational waves from the Cosmic Microwave Background}.
\newblock {\em JCAP}, 10:002, 2020.

\bibitem{Kite:2020uix}
Thomas Kite, Andrea Ravenni, Subodh~P. Patil, and Jens Chluba.
\newblock {Bridging the gap: spectral distortions meet gravitational waves}.
\newblock {\em Mon. Not. Roy. Astron. Soc.}, 505(3):4396--4405, 2021.

\bibitem{Lasky:2015lej}
Paul~D. Lasky et~al.
\newblock {Gravitational-wave cosmology across 29 decades in frequency}.
\newblock {\em Phys. Rev. X}, 6(1):011035, 2016.

\bibitem{LISA:2024hlh}
Monica Colpi et~al.
\newblock {LISA Definition Study Report}.
\newblock 2 2024.

\bibitem{KAGRA:2021kbb}
R.~Abbott et~al.
\newblock {Upper limits on the isotropic gravitational-wave background from Advanced LIGO and Advanced Virgo\textquoteright{}s third observing run}.
\newblock {\em Phys. Rev. D}, 104(2):022004, 2021.

\bibitem{Branchesi:2023mws}
Marica Branchesi et~al.
\newblock {Science with the Einstein Telescope: a comparison of different designs}.
\newblock {\em JCAP}, 07:068, 2023.

\bibitem{2017PhRvD..95f3002C}
Aaron~S. {Chou}, Richard {Gustafson}, Craig {Hogan}, Brittany {Kamai}, Ohkyung {Kwon}, Robert {Lanza}, Shane~L. {Larson}, Lee {McCuller}, Stephan~S. {Meyer}, Jonathan {Richardson}, Chris {Stoughton}, Raymond {Tomlin}, Rainer {Weiss}, and {Holometer Collaboration}.
\newblock {MHz gravitational wave constraints with decameter Michelson interferometers}.
\newblock {\em Phys. Rev. D}, 95(6):063002, March 2017.

\bibitem{Goryachev:2021zzn}
Maxim Goryachev, William~M. Campbell, Ik~Siong Heng, Serge Galliou, Eugene~N. Ivanov, and Michael~E. Tobar.
\newblock {Rare Events Detected with a Bulk Acoustic Wave High Frequency Gravitational Wave Antenna}.
\newblock {\em Phys. Rev. Lett.}, 127(7):071102, 2021.

\bibitem{Sesana:2019vho}
Alberto Sesana et~al.
\newblock {Unveiling the gravitational universe at $\mu$-Hz frequencies}.
\newblock {\em Exper. Astron.}, 51(3):1333--1383, 2021.

\bibitem{Ellis:2023owy}
John Ellis, Malcolm Fairbairn, Gert H\"utsi, Martti Raidal, Juan Urrutia, Ville Vaskonen, and Hardi Veerm\"ae.
\newblock {Prospects for future binary black hole gravitational wave studies in light of PTA measurements}.
\newblock {\em Astron. Astrophys.}, 676:A38, 2023.

\bibitem{Blas:2021mqw}
Diego Blas and Alexander~C. Jenkins.
\newblock {Bridging the \ensuremath{\mu}Hz Gap in the Gravitational-Wave Landscape with Binary Resonances}.
\newblock {\em Phys. Rev. Lett.}, 128(10):101103, 2022.

\bibitem{Foster:2025nzf}
Joshua~W. Foster, Diego Blas, Adrien Bourgoin, Aurelien Hees, M\'\i{}riam Herrero-Valea, Alexander~C. Jenkins, and Xiao Xue.
\newblock {Discovering $\mu$Hz gravitational waves and ultra-light dark matter with binary resonances}.
\newblock 4 2025.

\bibitem{Hui:2012yp}
Lam Hui, Sean~T. McWilliams, and I-Sheng Yang.
\newblock {Binary systems as resonance detectors for gravitational waves}.
\newblock {\em Phys. Rev. D}, 87(8):084009, 2013.

\bibitem{Blas:2021mpc}
Diego Blas and Alexander~C. Jenkins.
\newblock {Detecting stochastic gravitational waves with binary resonance}.
\newblock {\em Phys. Rev. D}, 105(6):064021, 2022.

\bibitem{Foster:2025csl}
Joshua~W. Foster, Diego Blas, Adrien Bourgoin, Aurelien Hees, M\'\i{}riam Herrero-Valea, Alexander~C. Jenkins, and Xiao Xue.
\newblock {Prospects for gravitational wave and ultra-light dark matter detection with binary resonances beyond the secular approximation}.
\newblock 4 2025.

\bibitem{2014PhRvD..89h2002L}
David~M. {Lucchesi} and Roberto {Peron}.
\newblock {LAGEOS II pericenter general relativistic precession (1993-2005): Error budget and constraints in gravitational physics}.
\newblock {\em Phys. Rev. D}, 89(8):082002, April 2014.

\bibitem{Lucchesi:2010zzb}
David~M. Lucchesi and Roberto Peron.
\newblock {Accurate Measurement in the Field of the Earth of the General-Relativistic Precession of the LAGEOS II Pericenter and New Constraints on Non-Newtonian Gravity}.
\newblock {\em Phys. Rev. Lett.}, 105:231103, 2010.

\bibitem{Blas:2016ddr}
Diego Blas, Diana~Lopez Nacir, and Sergey Sibiryakov.
\newblock {Ultralight Dark Matter Resonates with Binary Pulsars}.
\newblock {\em Phys. Rev. Lett.}, 118(26):261102, 2017.

\bibitem{Blas:2019hxz}
Diego Blas, Diana L\'opez~Nacir, and Sergey Sibiryakov.
\newblock {Secular effects of ultralight dark matter on binary pulsars}.
\newblock {\em Phys. Rev. D}, 101(6):063016, 2020.

\bibitem{Bertotti:2003rm}
B.~Bertotti, L.~Iess, and P.~Tortora.
\newblock {A test of general relativity using radio links with the Cassini spacecraft}.
\newblock {\em Nature}, 425:374--376, 2003.

\bibitem{Champion:2008ge}
D.~J. Champion et~al.
\newblock {An Eccentric Binary Millisecond Pulsar in the Galactic Plane}.
\newblock {\em Science}, 320:1309--1312, 2008.

\bibitem{2011MNRAS.412.2763F}
P.~C.~C. {Freire}, C.~G. {Bassa}, N.~{Wex}, I.~H. {Stairs}, D.~J. {Champion}, S.~M. {Ransom}, P.~{Lazarus}, V.~M. {Kaspi}, J.~W.~T. {Hessels}, M.~{Kramer}, J.~M. {Cordes}, J.~P.~W. {Verbiest}, P.~{Podsiadlowski}, D.~J. {Nice}, J.~S. {Deneva}, D.~R. {Lorimer}, B.~W. {Stappers}, M.~A. {McLaughlin}, and F.~{Camilo}.
\newblock {On the nature and evolution of the unique binary pulsar J1903+0327}.
\newblock {\em Monthly Notices of the RAS}, 412(4):2763--2780, April 2011.

\bibitem{Weisberg:2004hi}
Joel~M. Weisberg and Joseph~H. Taylor.
\newblock {Relativistic binary pulsar B1913+16: Thirty years of observations and analysis}.
\newblock {\em ASP Conf. Ser.}, 328:25, 2005.

\bibitem{Roy:2025iug}
Soumen Roy, Bruno Bertrand, and Justin Janquart.
\newblock {Probing gravitational waves using GNSS constellations}.
\newblock In {\em {59th Rencontres de Moriond on Gravitation}: {Moriond 2025 Gravitation}}, 5 2025.

\end{thebibliography}

\end{document}